\documentclass[manuscript]{aastex}

\slugcomment{To Appear in ApJL March 10 2006}

\shorttitle{High-velocity Dispersion Molecular Hydrogen in Stephan's Quintet}
\shortauthors{Appleton et al.}

\begin{document}


\title{Powerful High Velocity-dispersion Molecular Hydrogen associated with an Intergalactic Shock Wave in Stephan's Quintet}


\author{P. N. Appleton\altaffilmark{1}, Kevin C. Xu\altaffilmark{2}, William Reach\altaffilmark{1}, Michael A. Dopita\altaffilmark{3}, Y. Gao\altaffilmark{4}, N. Lu\altaffilmark{1}, C. C. Popescu\altaffilmark{5}, J. W. Sulentic\altaffilmark{6}, R. J. Tuffs\altaffilmark{5} AND M. S. Yun\altaffilmark{7}  }
\altaffiltext{1}{Spitzer Science Center, California Institute of Technology, 1200 E. California Blvd., Pasadena CA 91125}
\altaffiltext{2}{IPAC, California Institute of Technology, 1200 E. California Blvd., Pasadena CA 91125}
\altaffiltext{3}{Research School of Astronomy and Astrophysics, The Australian National University, Cotter Road, Weston Creek, ACT 2611, Australia}
\altaffiltext{4}{Purple Mountain Observatory, Chinese Academy of Sciences, 2 West Beijing Road, Nanjing 210008, China}
\altaffiltext{5}{Max Planck Institute f\"ur Kernphysik, Saupfercheckweg 1, 69117 Heidelberg, Germany}  
\altaffiltext{6}{Department of Physics and Astronomy, University of Alabama, Tuscaloosa, AL 35487}
\altaffiltext{7}{Department of Astronomy, University of Massachusetts, Amherst, MA 01003}



\begin{abstract}
We present the discovery of strong mid-infrared emission lines of
molecular hydrogen of apparently high velocity dispersion ($\sim$870
km s$^{-1}$) originating from a group-wide shock wave in Stephans
Quintet.  These {\it Spitzer Space Telescope} observations reveal
emission lines of molecular hydrogen and little else. This is the
first time an almost pure H$_2$ line spectrum has been seen in an
extragalactic object. Along with the absence of PAH-dust features and
very low excitation ionized gas tracers, the spectra resemble shocked
gas seen in Galactic supernova remnants, but on a vast scale. The
molecular emission extends over 24 kpc along the X-ray emitting
shock-front, but has ten times the surface luminosity as the soft
X-rays, and about one-third the surface luminosity of the IR
continuum.  We suggest that the powerful H$_2$ emission is generated
by the shock wave caused when a high-velocity intruder galaxy collides
with filaments of gas in the galaxy group. Our observations suggest a
close connection between galaxy-scale shock waves and strong broad
H$_2$ emission lines, like those seen in the spectra of Ultraluminous
Infrared Galaxies where high-speed collisions between galaxy disks are
common.
\end{abstract}


\keywords{galaxies: interactions, intergalactic medium, individual (NGC 7318b), evolution.}



\section{Introduction} Stephan's Quintet (hereafter SQ) is a system of
four strongly interacting galaxies in a compact group, and a likely
foreground galaxy seen in projection against them \citep{all70}.  One
of the most remarkable aspects of SQ is a $\sim$40 kpc long
non-thermal radio continuum structure (see Fig.1) lying in
intergalactic space between the galaxies \citep{all72,van81}. The same
structure is also seen in X-rays, and has been shown to be consistent
with a large-scale shock wave, based on the optical spectroscopy
\citep{pie97,tri05,xu03}. It is probable \citep{sul01} that the shock
wave has formed because a high-velocity ``intruder'' galaxy,
\objectname{NGC 7318b}, is colliding with the inter-group-medium (IGM)
located within the main group. We assume here a group distance of 94
Mpc, assuming a Hubble constant of 70 km s$^{-1}$ Mpc$^{-2}$ and a
systemic velocity for the group as a whole of 6600 km s$^{-1}$.

\section{Observations}

We used the Infrared Spectrograph \citep{jh04} on board the {\it
Spitzer Space Telescope}\footnote{This work is based on observations
made with the {\it Spitzer Space telescope} which is operated by the
Jet Propulsion Laboratory, California Institute of Technology, under
NASA contract 1407.} to take mid-IR spectra at the peak in the
IR/radio emission from the shocked region. Observations were made on
November 17 and

\includegraphics[scale=1.0,angle=0]{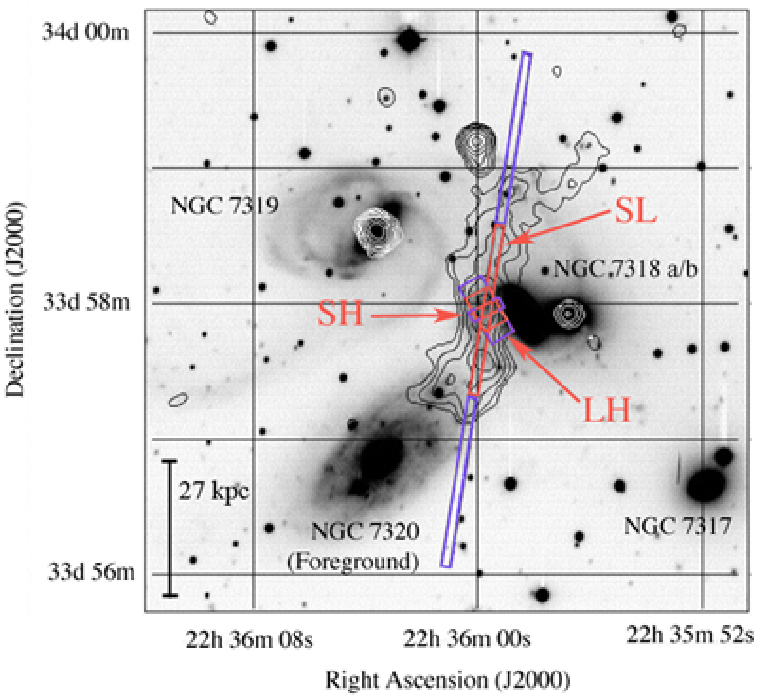}

{\scriptsize Fig.1 Stephan's Quintet showing the positioning of the IRS slits
centered on $\alpha$(J2000) = 22h 35m 59.57s, $\delta$(J2000) = +33d
58m 1.8s. Contours are VLA 1.4 GHz radio emission superimposed on an
R-band optical image\citep{xu03}. The X-ray emission (not shown here)
follows closely that of the radio. Only the central portion of the
slits (red boxes) were common to the two separate observing (nod)
positions made in each IRS module slit (the purple shows the full
coverage). Short-Low = SL = 57 x 3.6 arcsecs$^2$, Short-High = SH =
11.3 x 4.7 arcsec$^2$, and Long-High = LH = 22.3 x 11.1
arcsecs$^2$. The scale bar assumes a distance to the group of 94 Mpc.}

December 8 2004 using the Short-Low (SL), Short-Hi
(SH) and Long-Hi (LH) modules of the spectrograph covering the
wavelength range $\lambda$5.3-14.0$\mu$m, $\lambda$10.0-19.5$\mu$m and
$\lambda$18.8-37.2$\mu$m respectively. The raw data were processed
through the SSC IRS S11-science pipeline to create 2-d flat-fielded
images of the spectral orders. These data were further processed to
remove the effects of the so-called rogue pixels which were not
corrected by the pipeline. These pixels have erratic high
dark-currents which vary with time, and were replaced using a
simple pixel interpolation scheme. Flux and wavelength calibration was
performed using the standard calibration methods \citep{dec04}. Final
extraction of the 1-d spectra was made using the SSC software SPICE,
and line fluxes were measured using SMART.

\includegraphics[scale=0.5,angle=0]{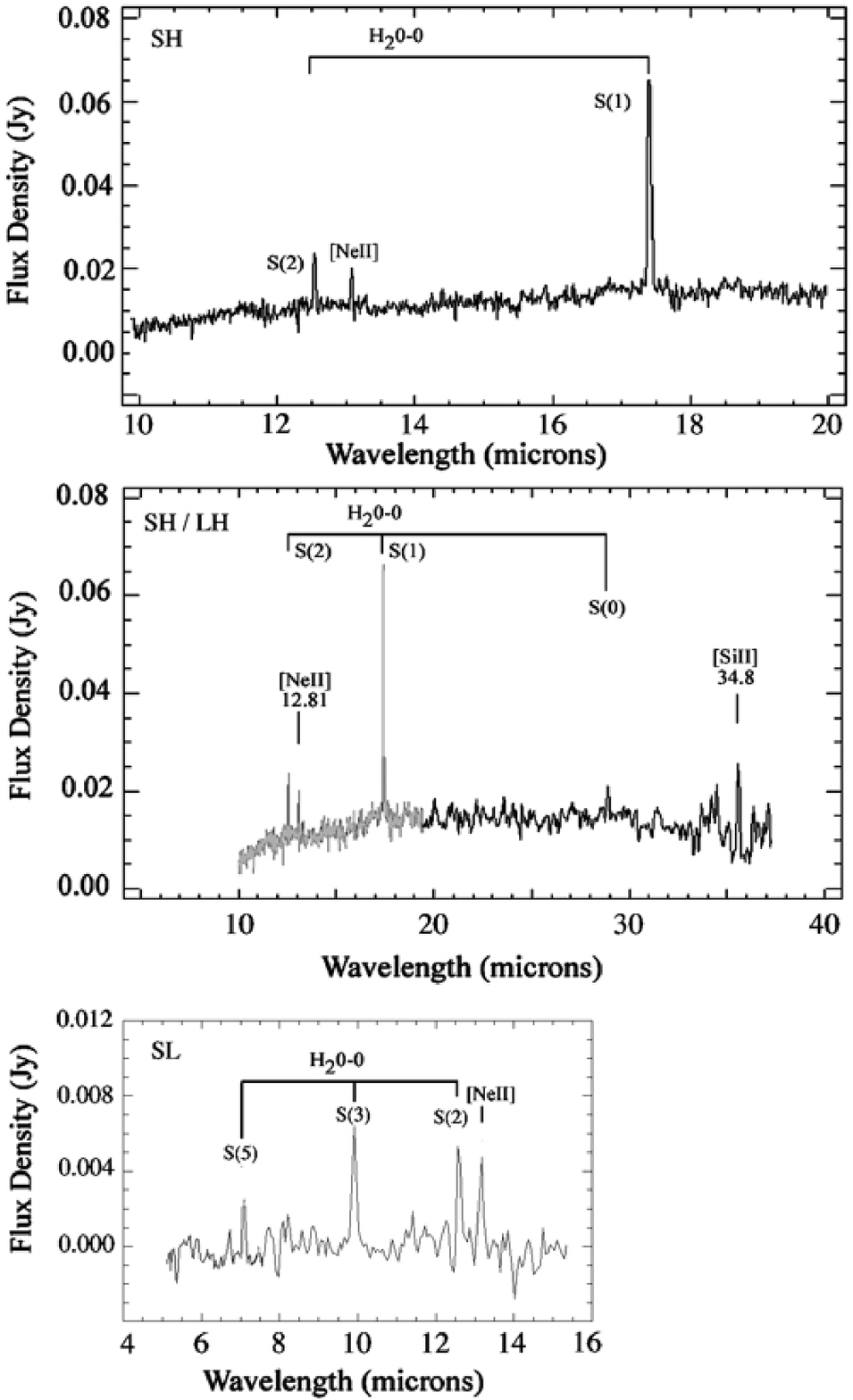}

{\scriptsize Fig.2 a) The IRS SH spectrum of the brightest radio/IR point in the
shock-front.  b) the combined SH (grey) and LH (black) spectrum of the
same target region assuming that the continuum can be used to match
the two spectra at 20$\mu$m, and c) the SL spectrum at the same
position. The SL2 (grey) and SL1 (black). Note that there is no
evidence for PAH emission.}

%% In this section, we use  the \subsection command to set off
%% a subsection.  \footnote is used to insert a footnote to the text.


%% This section also includes several of the displayed math environments
%% mentioned in the Author Guide.

\section{Results} Figure 2a, b and c show the spectra extracted in the
region where all three slits overlap.  Except for the atomic lines of
[NeII]$\lambda$12.8$\mu$m and [SiII]$\lambda$34.8$\mu$m, all are
rotational transitions of the ground vibrational states of molecular
hydrogen (the 0-0 S(0), S(1), S(2), S(3) and S(5) lines).  To explore
the excitation of the observed H$_2$ gas we need to compare the line
fluxes from the IRS slits which have different projected sizes on the
sky: requiring that we make some assumptions about the spatial
distribution of the molecules over the scale of the slits. From the SL
slit oriented along the shock structure, we found that the H$_2$
emission was extended on the scale of 52 arcsecs (24 kpc), declining
slowly from a small peak at the pointing centre.  We therefore
consider three possible distributions: a) the H$_2$ originates in a
point source--an unrealistic limiting case, b) the H$_2$ is extended
with constant surface brightness over the slits, and c) a
``preferred'' case based on the observed SL distribution, assuming it
has a similar spatial width to the radio emission. Table 1
provides a list of the lines and their measured fluxes, including
upper limits for some key undetected lines.

Figure 3 shows the excitation diagram for H$_2$, (N/g) versus the
upper-level energy, for the 5 detected and one undetected hydrogen
transitions. N is the molecular column density, g is the statistical
weight for that transition.  The non-linear decline of log (N/g) with
upper-level energy is commonly seen in shocks within the Galaxy, as
well as in external galaxies \citep{lut03,rig02}, and is an indication
that no single temperature LTE model fits these data. There are no
obvious deviations from the assumed ortho/para ratio = 3 visible in
these plots. We fit a multi-temperature model through the
``preferred'' data points, constraining the warm gas by the S(0)/S(2)
[para] ratio, and allowing the higher order S(3)/S(5) [ortho]
transitions to provide a very rough guide to the temperature of a
hotter component. Our data are consistent with a warm T~= ~185$\pm$30K
component (H$_2$ column density 2.06 $\times$ 10$^{20}$ cm$^{-2}$) and
T~=~675$\pm$80K (column 1.54 $\times$ 10$^{18}$ cm$^{-2}$) for the
hotter components.  The choice of a 2-component temperature model
(especially the hotter component) is quite arbitrary: in reality (if
our interpretation is correct that the emission arises in a highly
turbulent medium) it is likely to be the sum of a continuous
distribution of multi-temperature components. However, the coolest
component dominates the hydrogen column density estimates given here.
We note that a multi-phase medium could also lead to a similar excitation
pattern.  

\includegraphics[scale=0.4,angle=0]{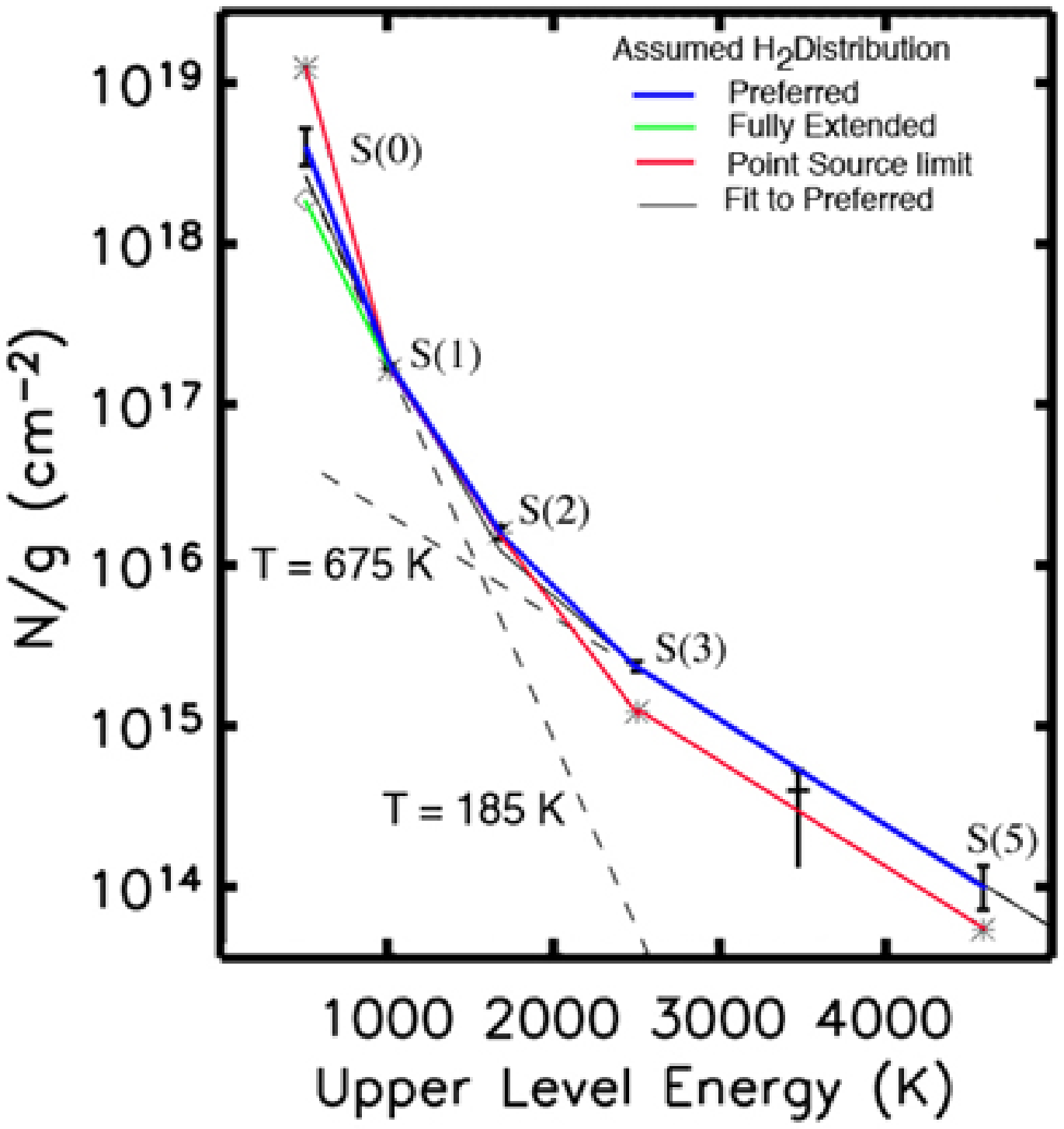}

{\scriptsize The excitation diagram for H$_2$.  The colored lines
represent three different assumed spatial distributions of the H$_2$
across the LH and SH slits. For SL, we scale the spectrum to that of
SH using the 0-0 S(2) line which is detected in common.  The blue line
(and error bars) represents the ``preferred'' distribution based on an
extrapolation from the H$_2$ distribution seen in the SL slit (see
text), the green line and red show the limiting cases of an infinite
and point source distribution respectively.  The dotted lines
represent a rough model for the ``preferred'' distribution which
assumes a two-temperature fit to the points. The solid line represents
the sum of these components.}

The total H$_2$ line luminosity, accounting for an extra 40\% in
undetected lines, is 8.4 $\times$ 10$^{40}$ ergs s$^{-1}$ from the SH
slit area alone. The corresponding H$_2$ mass in these warm/hot
components seen in the SH slit is 3.4 $\times$ 10$^7$ M$_{\odot}$
($\Sigma$$_{H_2}$ = 3.1 M$_{\odot}$ pc$^{-2}$). The H$_2$ surface
luminosity of 2.0 L$_{\odot}$ pc$^{-2}$, is only a factor of three
less than the IR continuum surface luminosity of 5.4 L$_{\odot}$
pc$^{-2}$ (= 2.5 $\times$ 10$^{41}$ ergs s$^{-1}$) based on an
extrapolation from unpublished 70$\mu$m Spitzer data by one of us
(RJT--and consistent with earlier ISO data from Xu et al. 2003). The
corresponding H$_2$ lines and FIR continuum surface brightnesses are
10$\times$ and 26$\times$ the X-ray (0.5--1.5~kev) surface brightness
(0.21 L$_{\odot}$ pc$^{-2}$) respectively. The dominance of the FIR
continuum emission over the X-ray emission confirms that the shocked
gas is cooling primarily through inelastic collisions of the ions and
electrons with grains, as proposed by (Xu et al. 2003--who derived a
cooling timescale of $\sim$2~Myrs), and may explain the unexpectedly
cool post-shock X-ray gas seen in recent XMM observations
\citep{tri05}. However, the fact that a substantial minority of the
total cooling also proceeds through H$_2$ lines comes as a surprise.

A remarkable feature of the Spitzer observations (Fig.4) is the
discovery that the H$_2$ line is extremely broad and resolved
spectrally, even with the relative low resolution of the IRS
(R=600). Assuming a Gaussian decomposition from the instrument
profile, we estimate the intrinsic width of the 0-0~S(1) line to be
870 $\pm$60 km s$^{-1}$: exceeding the largest known H$_2$ line-width
known to date (e. g. the ULIRG NGC 6240; $\Delta$V $\sim$ 680 km
s$^{-1}$) \citep{lut03,arm05}. Because H$_2$ is a fragile molecule
which can easily be destroyed in 50 km s$^{-1}$ J-shocks
\citep{hol80,hol89}, such a wide, potentially intrinsic, spread in
molecular cloud velocities is extremely unusual. 

The line-width is comparable with the collision-velocity ($\Delta$V
$\sim$500-600 km s$^{-1}$) of NGC 7318b with either NGC 7319 or an
inter-galactic gas filament (Allen \& Hartsuiker 1972; van der Hulst
1981; Williams et al. 2002; Sulentic et al. 2001; Trinchieri et
al. 2003; Trinchieri et al. 2005), suggesting that the H$_2$ emission
is intimately linked to the formation of the radio and X-ray shock
structure.

\includegraphics[scale=0.4,angle=0]{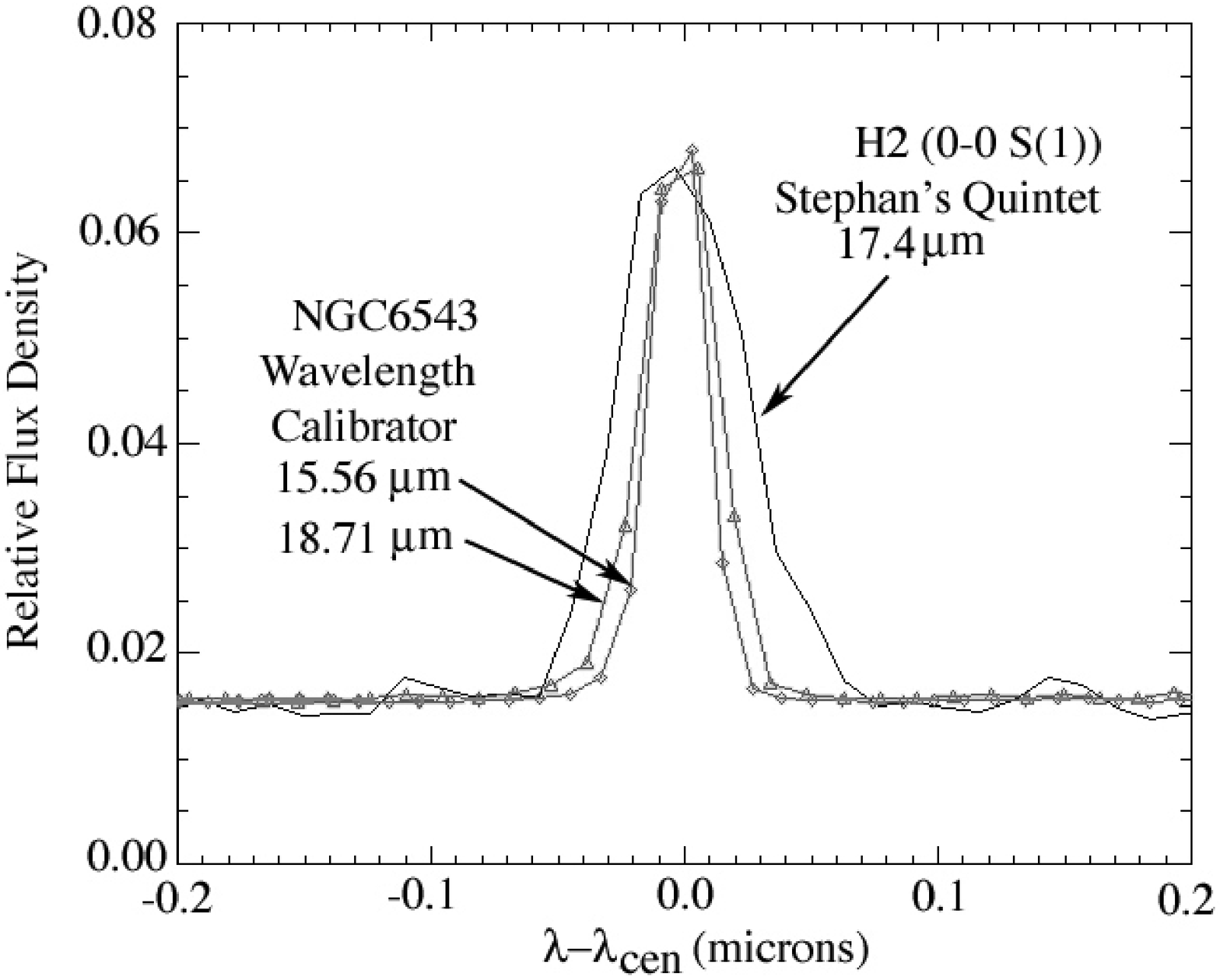}

{\scriptsize Fig.4 The 0-0 S(1)17.0$\mu$m molecular hydrogen line
(detected with a signal to noise ratio of 33) is clearly resolved
(intrinsic FWHM = 870$\pm$60 km s$^{-1}$) by the IRS (black) as
compared with two unresolved spectral lines (R = 600) in the planetary
nebula NGC 6543 at 15.6 and 18.7$\mu$m which bracket the H$_2$ line
(grey). Similar broad widths are found for the 0-0~S(0) and 0-0~S(2)
lines. The width of the [NeII]$\lambda$12.8$\mu$m line is significantly
less broad (intrinsic FWHM = 600$\pm$80 km s$^{-1}$). $\lambda$$_{cen}$ is the wavelength of the line center for the three lines shown in order to allow their widths to be compared. }

Although the IRS does not have the spectral resolution to determine
whether the apparently broad line is made up of a number of narrow
line components blended together along the line of sight, or whether
the line is truly broad, there are some pieces of evidence that favor
the latter. These are, 1) broad ($\Delta$V = 1000 km s$^{-1}$) optical
[OI]$\lambda$6300 line emission with excitation properties of shocked
gas is seen at our observing position \citep{xu03}, whereas at other
positions much further from the peak of the X-ray/IR ridge narrower
lines are seen, (2) Extremely high speed shocks are highly
dissociative, and it is unlikely that pre-existing H$_2$, nor even
more fragile PAH molecules, could survive a $>$500 km s$^{-1}$
collision with the intruder \citep{hol80,hol89}. Furthermore, the
H$_2$ line luminosity we see from the region ($\sim$2 L$_{\odot}$
pc$^{-2}$) is a significant fraction of the predicted
\citep{dop95,dop96} mechanical luminosity density in the shock (160
$\times$ the H$\alpha$ surface density, $\sigma{_{H\alpha}}$ = 160 x
0.13 L$_{\odot}$ pc$^{-2}$ = 21 L$_{\odot}$ pc$^{-2}$). There would
appear to be plenty of mechanical luminosity available to excite the
H$_2$ molecules if they reformed in post-shocked gas.

If we assume a 500 km s$^{-1}$ shock is driven into the IGM, and
second similar shock into the intruder (assuming their densities are
similar), this will create an unstable turbulent ($\Delta$V $\sim$
1000 km s$^{-1}$) cooling layer where the flows mix. If the gas can
cool quickly enough, H$_2$ recombination would occur in that layer,
and this would have the necessary properties to explain our
observations--including the high velocity dispersion from turbulence.
The cooling timescale~\citep{dop03} for the hot X-ray gas in each
shock wave, $\tau$$_{cool}$ (Myrs) = 0.24 $(V_{500})^{4.4} /Z n $,
where V$_{500}$ is the speed of the shock in units of 500 km s$^{-1}$,
Z is the metallicity in solar units and $n$ is the pre-shock density
(we assume $n$ = 0.027 cm$^{-3}$ , from X-ray observations
\citep{tri03}).  For $V_{500}$ $\sim$ 1, $Z$ = 1, $\tau$$_{cool}$ =
9.5 Myrs, during which time the shock wave would have travelled $\sim$
5kpc = 10 arcsecs (or twice this for the double shock). In reality,
the cooling layer would be much sharper, since this analysis excludes
the effects of IR-cooling by dust which is known to be very
significant \citep{xu03}.  This argument demonstrates that the
post-shocked gas layer would cool fast enough to begin forming H$_2$
within the viewing zone of the IRS slits even for a mainly transverse
geometry for the shock propagation as projected on the sky, The
observed H$_2$ column of the cooler component ($\sim$2 $\times$
10$^{20}$ cm$^{-3}$) could easily be built up over a scale of a few
kpc from the available pre-shocked gas mass once the temperature of
the gas had cooled sufficiently for H$_2$ to form.

One challenge for the post-shock origin of the H$_2$ is to avoid
putting too much of the shock dissipation energy into the X-ray
component. Usually in high-speed shocks, X-ray dissipation dominates
the cooling in the T= 10$^6$ to 10$^4$ K regime, before H$_2$ molecules
can form: not the case here since H$_2$ line emission exceeds the
X-rays by a factor of ten. Part of the answer may come from the even
stronger IR continuum seen at longer wavelengths (Xu et al
2003). These authors argued that the principal coolant in the shock is
from large grains which efficiently cool the shock through far-IR
emission.  The H$_2$ emission from the post-shocked material could be
additionally boosted relative to the X-rays by invoking a glancing
collision between the intruder galaxy and the IGM--which would create
an oblique-shock geometry rather than a plane-parallel one. In oblique
shocks, much more energy is carried in bulk transverse motions (hence
reducing X-ray dissipation), and the shock is significantly weakened
(e. g. Dopita 2003), while still allowing a fraction of the mechanical
energy to cascade down to smaller and smaller scales (as in the
Galactic ISM e. g. McKee, Chernoff \& Hollenbach 1984; Elmegreen \&
Scalo 2004), where shock-speeds would be lower and molecules could
re-form in the post-shocked medium.  To test these ideas will require
detailed 3-d hydrodynamic models beyond the scope of this
observational paper.

An alternative explanation for the H$_2$ emission is to postulate that
the emission comes from a 500 km/s shock overrunning a clumpy
pre-shocked medium. This has the advantage that dense clouds (embedded
in a more diffuse medium) will experience much slower shocks and are
more likely to cool through line emission than X-rays, whereas the FIR
emission could come from grains emitting over the entire diffuse
medium--thus partly explaining the anomalous H$_2$ line to IR
continuum ratio.  The large velocity dispersion could arise because
the shock would eventually accelerate clouds to different velocities
within the shock, or because of multiple velocity components being
present in the pre-shocked gas.  The problem with this picture is that
we might expect to see PAH emission from the precursor ISM which is
not observed. We note that if the H$\alpha$ emission seen from the
shocked regions was scaled to an equivalent PAH line strength typical
of those seen in the largest of the M51 HII regions (Calzetti et
al. 2005), we should have easily ($>$10-sigma) detected the signal
even in the case of no extinction correction for the H$\alpha$. Of
course the ISM of any precursor material is likely to contribute
only a small factor to the H$\alpha$ emission observed, but this
provides an upper limit. The shock is also likely to be radiative,
which could also help to destroy PAH molecules, whereas the more
refactory grains must have been able to survive the shock. This model
still has to be capable of predicting the large ratio of H$_2$ line
luminosity relative to the IR continuum luminosity--implying a lot of
energy has to be somehow channeled into mechanical energy.  Further
modeling of these scenario will be required in order to determine
whether it can explain all the available data on the shock.

Future H$_2$ observations with much higher velocity and spatial
resolution may allow us to explore further the nature of the emission
mechanism that is generating such large H$_2$ line flux. There are
hints of large-scale clumpyness in the H$_2$ distribution as
seen by SL along the shock front, and studying the velocity and
spatial inhomogenities within the shock would be highly beneficial.
   
\section{Implications for ULIRGs}

The discovery of strong, high velocity-dispersion H$_2$ emission in a
large-scale group-wide shock wave provides support for the idea
\citep{rie85}, that shock waves are primarily responsible for the
strong H$_2$ emission lines seen in many Ultraluminous Infrared
Galaxies (ULIRGs). Indeed our observations suggest L(H$_2$) $>$
10$^{41}$ ergs s$^{-1}$ over the whole shock structure--comparable to
the H$_2$ luminosity of Arp 220, but a factor of ten less than NGC
6240 \citep{arm05}.  An alternative idea for NGC 6240 is that X-rays
heating the gas, rather than shocks, is the dominant mechanism for the
IR lines in ULIRGs \citep{dra90}. However, in addition to the low X-ray
surface density compared with the H$_2$ lines, our spectra of
SQ, show little evidence of the
H$_3$$^+$$\lambda$16.33$\mu$m line: a key diagnostic of this model.  
Although the geometry of the shocks are unlikely to be
identical to that suggested by the SQ observations,
there is little doubt that large-scale shocks must be present in
ULIRGs, where high velocity streams of molecular gas must
collide--especially in the early stages of a merger.


%% Included in this acknowledgments section are examples of the
%% AASTeX hypertext markup commands. Use \url without the optional [HREF]
%% argument when you want to print the url directly in the text. Otherwise,
%% use either \url or \anchor, with the HREF as the first argument and the
%% text to be printed in the second.

\acknowledgments The authors thank A. Noriega-Crespo, L. Armus,
D. Shupe, J. Ingalls, G. Helou (SSC), V. Charmandaris (U. of Crete)
and J. Houck (Cornell U.) for comments, and the referee, D. Hollenbach, 
for valuable suggestions.

\clearpage

\begin{deluxetable}{lcccc}
\tabletypesize{\scriptsize}
\tablecolumns{5}
\tablewidth{0pt}
\tablecaption{Line Fluxes Scaled to the Equivalent SH-slit Aperture}
\tablehead{ 
   \colhead{Spectral} &
   \colhead{IRS Module} &
   \colhead{Flux} &
   \colhead{Flux} &
   \colhead{Flux} \\
   \colhead{Feature} &
   \colhead{} &
   \colhead{ $\times$ 10$^{-22}$ W cm$^{-2}$} &
   \colhead{ $\times$ 10$^{-22}$ W cm$^{-2}$} &
   \colhead{ $\times$ 10$^{-22}$ W cm$^{-2}$} \\
   \colhead{} &
   \colhead{} &
   \colhead{ (Uniform)\tablenotemark{a} } &
   \colhead{(Point Source Limit)\tablenotemark{a} } &
   \colhead{ (Preferred)\tablenotemark{a} }
  }
\startdata 
H$_2$ 0-0 S(0) & Long High & 1.9$\pm$0.5 & 12.6$\pm$3.3 & 4.2$\pm$1.1 \\ 
H$_2$ 0-0 S(1) & Short High & 19.80$\pm$0.57 & 19.80$\pm$0.57 & 19.80$\pm$0.57 \\ 
H$_2$ 0-0 S(2) & Short High & 6.41$\pm$0.58 & 6.41$\pm$0.58 & 6.41$\pm$d0.58 \\ 
H$_2$ 0-0 S(3) & Short Low & 15.6$\pm$1.1 & 8.1$\pm$1.1 & 15.6$\pm$1.1 \\ 
H$_2$ 0-0 S(4) & Short Low & $<$ 3.4$\pm$1.1 & $<$ 1.8$\pm$1.1 & $<$ 3.4$\pm$1.1 \\ 
H$_2$ 0-0 S(5) & Short Low & 7.8$\pm$2.4 & 4.1$\pm$2.4 & 7.8$\pm$2.4 \\ 
$[NeII]$12.81$\mu$m & Short High & 3.49$\pm$0.64 & 3.49$\pm$0.64 & 3.49$\pm$0.64 \\
$[NeIII]$15.55$\mu$m & Short High &  $<$1.5$\pm$0.64\tablenotemark{b} & $<$1.5$\pm$0.64\tablenotemark{b} & $<$1.5
$\pm$0.64\tablenotemark{b} \\
H$_3$$^{+}$ 16.33$\mu$m & Short High & $<$1.5$\pm$0.64 & $<$1.5$\pm$0.64 & $<$1.5$\pm$0.64 \\
$[SiII]$34.8$\mu$m & Long-High & 3.5$\pm$0.9 & 23.7$\pm$5.7 & 7.9$\pm$1.9 \\
\enddata
\tablenotetext{a}{Assumed distribution over scale of Long-Hi module (see text)} 
\tablenotetext{b}{Line may be marginally detected.}
\end{deluxetable}

\end{document}